# LaMPost: Design and Evaluation of an AI-assisted Email Writing Prototype for Adults with Dyslexia

STEVEN M. GOODMAN, University of Washington, USA and Google Research, USA

ERIN BUEHLER, PATRICK CLARY, ANDY COENEN, AARON DONSBACH, TIFFANIE N. HORNE, MICHAL LAHAV, ROBERT MACDONALD, RAIN BREAW MICHAELS, AJIT NARAYANAN, MAHIMA PUSHKARNA, JOEL RILEY, ALEX SANTANA, LEI SHI, RACHEL SWEENEY, PHIL WEAVER, and ANN YUAN, Google, USA

MEREDITH RINGEL MORRIS, Google Research, USA

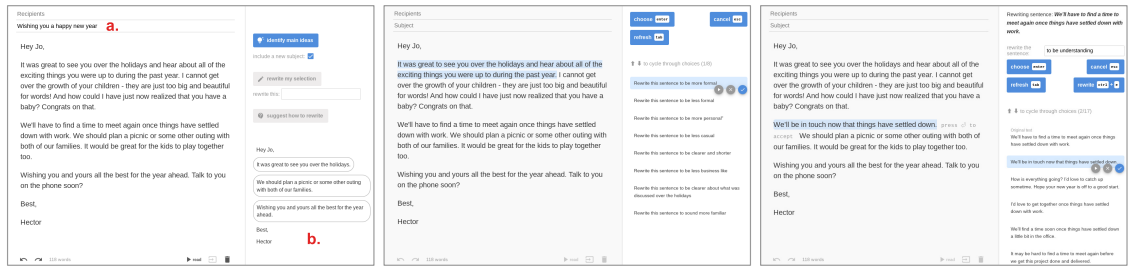

Fig. 1. The LaMPost interface. The system augments a typical browser-based email editor with three AI-powered features: (left) users can generate an outline of the email's main ideas (a) with the option for a related subject line (b); (center) users can generate suggestions for possible changes to a selected passage, and (right) users can generate rewritten text for a selected passage based on a human- or machine-written instruction.

Prior work has explored the writing challenges experienced by people with dyslexia, and the potential for new spelling, grammar, and word retrieval technologies to address these challenges. However, the capabilities for natural language generation demonstrated by the latest class of large language models (LLMs) highlight an opportunity to explore new forms of human-AI writing support tools. In this paper, we introduce *LaMPost*, a prototype email-writing interface that explores the potential for LLMs to power writing support tools that address the varied needs of people with dyslexia. LaMPost draws from our understanding of these needs and introduces novel AI-powered features for email-writing, including: outlining main ideas, generating a subject line, suggesting changes, rewriting a selection. We evaluated LaMPost with 19 adults with dyslexia, identifying many promising routes for further exploration (including the popularity of the "rewrite" and "subject line" features), but also finding that the current generation of LLMs may not surpass the accuracy and quality thresholds required to meet the needs of writers with dyslexia. Surprisingly, we found that participants' awareness of the AI had no effect on their perception of the system, nor on their feelings of autonomy, expression, and self-efficacy when writing emails. Our findings yield further insight into the benefits and drawbacks of using LLMs as writing support for adults with dyslexia and provide a foundation to build upon in future research.

## 1 INTRODUCTION

Dyslexia refers to a cluster of symptoms that result in challenges with word recognition, reading fluency, spelling, and writing that impacts up to 20% of the population [1, 32, 59]. While some adults with dyslexia may learn and adopt compensatory strategies for reading difficulties over time [39, 42, 44], the combination of reading, comprehension, and planning skills required to carry out writing tasks may lead to on-going difficulties [42, 55]. In addition to low-level obstacles such as spelling and grammar, writers with dyslexia report a variety of high-level challenges (*e.g.*. [13, 42, 47]), such as ordering and expressing their ideas, choosing language to match their desired tone, and writing with clarity and precision. To overcome these obstacles, they report using a variety of strategies—such as speech-to-text tools to dictate ideas, templates to match style, and revising feedback from friends and family—but these can add further complexity and time to their writing process [13, 46].

Prior work in accessibility has explored a number of approaches to overcome the reading challenges associated with dyslexia, such as experimenting with various forms of text presentation [19, 50]) and synonym substitution for complex words [51]. However, work targeting dyslexia's associated writing challenges has primarily focused on low-level interventions, including automatic suggestions to support word retrieval [37, 41, 49] and specialized spellcheck tools (*e.g.*, [37, 45, 52, 66]). AI-based efforts, when present, have continued this thread; for example, Wu et al. evaluated a dyslexia-tuned Neural Machine Translation model for spelling and grammar support on social media posts [66]. However, tools that can lend support to people with dyslexia for important high-level aspects of writing—such as organization, expression, and voice—are absent from accessibility literature. This gap highlights an opportunity to explore the potential for AI-powered writing support tools that use state-of-the-art neural language models.

Neural language models are neural networks that are trained to predict the next word in a sequence given the previous words. We use "large language models," or LLMs, to refer to the recent class of neural language models (*e.g.*, GPT-3 [7]) that have been trained using the Transformer neural architecture [64] and are capable of generating long passages of text that human evaluators perceive as human-written [15]. With *few-shot* learning to enable controllable text generation, LLMs hold potential to drive new technologies that bolster written expression [69]. This functionality may provide significant value to writers with dyslexia by alleviating common difficulties and simplifying their existing workflow, but questions arise over the correct approach for their implementation. For example, although automatic text generation could help some writers with dyslexia to conquer their "fear of the blank page" [47], machine-powered writing may raise concerns over the authors' control and autonomy in the writing process [27].

In this paper, we introduce LaMPost, an LLM-based prototype to support adults with dyslexia with writing emails. LaMPost implements LaMDA [60], an LLM for dialog applications, to augment a standard email editor with AI-powered outlining, subject generation, suggestion, and rewriting features. We evaluated LaMPost with 19 adult participants with dyslexia. Our findings indicate enthusiasm among this demographic for high-level writing support features, including rewriting passages in a particular tone or style (*e.g.*, "more formal", "more concise") and generating summative content such as subject lines based on an email's body. However, we also found that accuracy and quality issues in the current generation of LLMs present obstacles to a reliable and trustworthy writing-support experience. Further, effectively utilizing LLMs for writers with dyslexia may require HCI innovations to manage tradeoffs, such as autonomy *vs.* cognitive load and personalization *vs.* privacy. Knowledge that our writing-support tool contained AI did not have a significant effect on participants' perception of the system and written work. Our findings highlight opportunities and challenges of AI-assisted writing support for people with dyslexia and provide a foundation for future work as the capabilities of generative language models—and our understanding of their risks and trade-offs—mature.

## 2 RELATED WORK

Our research is informed by and builds upon: work on dyslexia and associated writing challenges, prior accessibility research with this population, and AI-assisted writing tools.

### 2.1 Writers with Dyslexia

Dyslexia is a multifaceted condition characterized by difficulties with word recognition, reading fluency, spelling, and/or writing [59]. According to van Schaik [63], the complete definition of dyslexia varies according to the lens studying it. Through a medical lens, dyslexia is defined as a cognitive deficiency that is associated with persistent difficulties with reading, spelling, short-term/working memory, and day-to-day organization (*e.g.*, [13, 24]). Through a lens of neurodiversity, however, dyslexia is defined by heightened spatial and perceptual abilities, interconnected and dynamic reasoning, and narrative and holistic thinking—alongside commonly defined deficits [4]. Critical scholars define dyslexia as a person's failure to meet the socially constructed expectations of timelines, literacy, and communication that are embedded in one's broader social and cultural context [17, 29]. While dyslexia impacts up to 20% of the population [1], structural disparities including gender, class, and race [42, 63] cause many cases to go undiagnosed—leading many to be unaware of the cause of their reading and writing difficulties.

While many dyslexia-related challenges remain into adulthood [8], individuals with dyslexia may learn and adopt compensatory strategies for reading difficulties over time [39, 42, 44]. Although reading may still prove challenging, some adults with dyslexia report that writing tasks tend to provide their greatest difficulties [42, 55]. Writing challenges are wide-ranging, but some are commonly reported [13, 18, 42, 47, 61]: on a high level (overall plan and structure), these include organizing and expressing one's thoughts, structuring and ordering ideas, and overcoming a "fear of the blank page" [47]. At a lower level (sentence and word), challenges can include word retrieval, sentence composition, appropriate tone and concision, grammar, spelling, punctuation, and proofreading. As with reading, writers with dyslexia may adopt strategies to assist in their writing—such as preferred spell-checkers, text-to-speech and dictation software, and support from friends and family—but these can add complexity and time to their writing process [13, 46].

In this paper, we contribute findings regarding the needs and challenges experienced by adults with dyslexia in email-writing, and we explore how these might be addressed with AI interventions.

### 2.2 Dyslexia & Accessible Technology Design

Because text readability is impacted by the visual display of text, researchers have explored how to alleviate reading challenges through text presentation, such as typography choices [50], word segmentation [3], background colors [53], and increased font size and margin space [19, 54]. To support reading comprehension, one promising approach involves text simplification [49]. Rello et al. [51] found promising results among readers with dyslexia when displaying basic synonyms alongside complex words, although readers struggled when a simpler word was substituted automatically. In a study of web searchers with dyslexia [41], participants sought pages utilizing multimedia whilst containing minimal visual clutter and large text blocks (preferring headings and bullets instead). In the design of our email-writing prototype, we draw from elements of this body of work to address usability challenges—including guidelines for text presentation and visual clutter—and explore text simplification as a form of writing support.

Work in accessibility targeting writers with dyslexia has primarily focused on low-level interventions, such as specialized spellcheck tools (*e.g.*, [37, 45, 52, 66]). While common spellcheck tools provide value to this population, specialized versions are motivated by the high occurrence of "real word" errors (*e.g.*, "hear" and "here") among people with dyslexia that most common tools cannot recognize [45, 52]. Text suggestions can also provide value when writing to overcome word retrieval difficulties [37, 41, 49]. PoliSpell [37] was an early attempt to design a spellcheck and autocomplete tool for people with dyslexia, but it was never evaluated. Wu et al. examined the experience of writers with dyslexia on social media, finding challenges associated with not only the writing task, but concerns over social self-presentation [55]—and used their results to build a dyslexia-tuned Neural Machine Translation model for spelling and grammar support on Facebook posts [66].

So far, accessibility researchers have not explored the high-level challenges associated with dyslexia during text construction, nor have they begun exploring AI-powered solutions to these challenges—a gap our work aims to address.

## 2.3 Large Language Models

The most recent class of large language models such as GPT-3 [7] demonstrate significant advances in natural language generation. At their core, these models have a simple API: given a string of text, known as a *prompt* [55], they return plausible continuations for that string. For example:

> **prompt:** A healthy lunch includes
> **language model:** fruits, vegetables, protein, and whole grains.

Prompts can also be written with exemplars of a desired response, such that the model ends up performing a specific task by continuing the text. In this way, they are capable of *few-shot learning* [7] which is shown to be more accurate than the *zero-shot* example above [71]. This example prompts the language model for a piece of clothing and an accessory to plan for the weather:

> **prompt:** When it's sunny, I need:
> shorts and sunscreen
> When it's raining, I need:
> rain boots and an umbrella
> When it's snowing, I need:
> **language model:** mittens and a shovel

Researchers have begun exploring many end-user facing applications powered by LLMs, including chatbots and conversational agents [68], code generation [2, 14, 70], creative writing [16, 28, 69], and even accessibility applications such as keystroke-saving abbreviations expansions to accelerate eye-gaze typing by users with motor disabilities [10].

While LLMs exhibit impressive performance on many tasks and have many potential applications, these models also have drawbacks. Of particular note is that such models risk generating factually incorrect, offensive, or stereotyped text since they are trained on content from the internet [5, 65]. "Memorization" (*e.g.*, regurgitating existing text rather than producing novel content) is also a risk of current LLMs [12]. The risks of erroneous or inappropriate output from LLMs carry additional ethical challenges when embedded in systems used by vulnerable audiences, such as users with dyslexia, who may experience challenges in interpreting the output's quality [25, 40]. Mitigating the risks associated with LLMs is an active area of research; for example, LaMDA (the LLM underpinning LaMPost) uses fine-tuning to improve model safety [60].

## 2.4 AI-Assisted Writing

Human-AI co-creation in the writing domain (for general audiences) has been widely studied, and applications such as Gmail's Smart Reply feature [35] have already been deployed to massive audiences. Buschek et. al. [9] explored the impact of multiple suggested text continuations when writing emails, finding benefits to ideation at the cost of writing speed. Gero et. al. [26, 27] studied automatic synonym and metaphor generation and found both features enabled greater expression during the writer's process, but questions arose over autonomy and ownership over the produced text [27]. Further questions arise over how the algorithms powering these systems should be presented to the user, as users' perceptions toward an AI system and desire to use it can be impacted by this choice of presentation [33, 36]. We explore these questions further in this work.

Wordcraft [16, 69] explored LLMs incorporated into the writing process: users collaborate with the model to write a story through a variety of operations—including infilling, elaboration, and rewriting—as well as open-ended dialog. The system augments a traditional text editor with a set of integrated LLM-powered controls driven by novel prompting techniques that enabled users to build their own custom controls, such as *"rewrite the text to be more melodramatic"*. In this paper, we adapt Wordcraft's approach to provide LLM-powered controls for writers with dyslexia, building an email editor with features for automatic outlining, subject generation, rewriting, and suggestions. We also focus specifically on how AI-assisted writing can support the needs of adults with dyslexia (rather than general audiences).

## 3 LAMPOST: AN LLM-POWERED PROTOTYPE FOR EMAIL WRITING SUPPORT

Prior to developing the LaMPost system, our interdisciplinary research team engaged in over a year of participatory research with the dyslexia community. This included participatory design sessions and workshops with partner organizations with expertise in reading disabilities[1], and culminated in a brainstorming workshop on AI-assisted writing support with experts in accessibility and dyslexia (including team members with lived experience of dyslexia and other visual processing and reading disabilities).

As a key step in our formative work, we conducted a 90-minute formative study to motivate the design of an AI-powered writing system. We recruited seven adults with dyslexia to join the two-part study for (1) individual interviews on writing practices and challenges, and (2) a group assessment of possible ideas for AI writing support. Individual interviews highlighted several challenges, such as: planning how to order ideas, expressing ideas in clear and concise wording, writing with appropriate tone, and finding proofreading help. Further, the group interviews highlighted an overall interest in AI writing support: revising feedback could help with clarity, verbosity, and tone; summarization could validate an intended meaning; and visual organization could help to order and structure ideas. However, concerns arose over users' capability to address AI feedback, maintaining autonomy and control over their work, and their privacy.

Informed by our formative inquiries and prior work [42, 46, 47], we built LaMPost, a web application and LLM-powered prototype for email writing support for writers with dyslexia (Figure 1). We chose email writing as a constrained—yet highly practical—use case to demonstrate and compare different approaches for building a text editor infused with generative writing support.

---

[1]Partner organizations included the British Dyslexia Association, Understood.org, Madras Dyslexia Association, and the Landmark School in Prides Crossing, MA.

In the following sections, we explain how LaMPost works and describe its design (including accessibility considerations to support users with dyslexia) and functionality (including the key motivations from our formative work for LaMPost's LLM-based features).

### 3.1 Email-Writing Support Through Few-Shot Learning

LaMPost is powered by LaMDA, a neural language model [60], and adapts the few-shot prompting methods introduced by the Wordcraft system [16, 69] (a LaMDA-powered tool for creative story-writing by general audiences). When the user selects one of LaMPost's operations, the system constructs a custom *few-shot learning* prompt and sends it to the language model.

Prompt performance is highly sensitive to word choice, formatting, and the content of the exemplars [71]. Writing effective prompts requires rigorous testing and iteration to achieve reliable and accurate responses from the model. When building LaMPost, we experimented with different prompting methods for several possible LLM-based features before settling on the three in our final system. We describe our iterative development process for LaMPost and reflect on lessons learned in Section 5.3.

### 3.2 LaMPost's Design and Functionality

LaMPost's interface consists of a main panel that resembles a standard email editor, including sections for the email's recipients, subject, and body; and "undo", "redo", and "clear" buttons near the bottom. A secondary panel on the right is reserved for three LLM-powered features: identifying main ideas (with the option to generate an email subject), rewriting a selection, and suggesting how to rewrite a selection. These three features were inspired by findings from our formative work with organizations and participants with dyslexia, as we describe in more detail below.

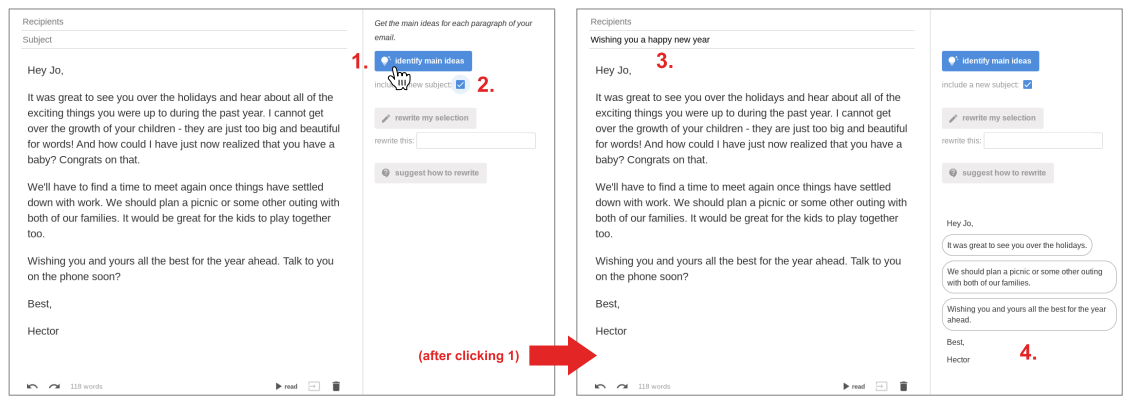

Fig. 2. **The Identify Main Ideas feature.** Users can click *'identify main ideas'* (1) and 'include a new subject' (2) to generate an outline of their email based on the main ideas of each paragraph (4) with a subject line generated on top (3). Hovering over each item of the outline highlights that respective paragraph in the email body.

*3.2.1 LLM Feature 1: Identify Main Ideas.* Users can generate a visual outline of their email with the main idea from each paragraph. Additionally, they can choose to generate a new subject line from this outline (Figure 2). This feature was motivated by feedback from our formative study with dyslexic adults, in which participants noted that to overcome difficulties with organizing ideas and making them understandable to readers, visual organization and automatic summarization were desired technological supports. The visual outline can make it easier to parse sections of a long text, while simplified content can make it easier to understand that text [49]. The option to generate a subject line allows users to ask the AI to simplify the content for them. By displaying the AI's interpretation of the email's salient points, we imagined this feature could also show users how the email's main ideas might be extracted by another reader.

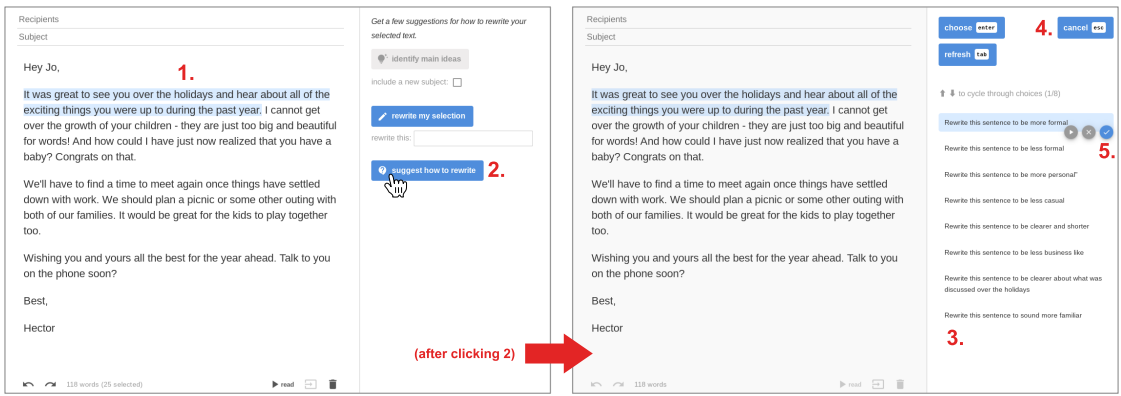

Fig. 3. **The Suggest Possible Changes feature.** A user can select a passage of text (1) and click 'suggest how to rewrite' (2). Several suggestions from the AI for changing the passage will populate in the right-hand panel (3). Users can choose to exit the operation and rewrite the text themselves (4), or choose to have individual suggestions 'read aloud', discarded, or to used as a prompt for the 'Rewrite My Selection' feature (5).

*3.2.2 LLM Feature 2: Suggest Possible Changes.* Users can select a word, phrase, or paragraph and ask the AI for suggestions on how to rewrite it (Figure 3). Participants in our formative study described feeling unsure about the kinds of adjustments needed for their writing, and were interested in automatic suggestions for high-level language characteristics like tone and clarity. Results from the LLM appear as several suggestions for changing the selected passage; for example, *"Rewrite this sentence to be less business-like"*. Users can take these suggestions into consideration to guide their own revisions, or use a preferred suggestion as a prompt to generate rewritten passages in a follow-up operation (described in the following section).

To implement this feature, the few-shot prompt included several examples containing: a passage from an email (*i.e.*, the user's selected), the full email to provide context, and a suggestion for improving the passage (the ideal response from the model). When users press the *'suggest how to rewrite'* button, their current selection and full email is appended to the end of the few-shot prompt and sent to the model; the model responds with several suggested changes to the passage for users to consider. We adapted the *meta-prompting* method from [69] to allow users to optionally use a suggestion as a precursor for the stand-alone rewriting feature described below.

*3.2.3 LLM Feature 3: Rewrite My Selection.* Users can select a word, phrase, or paragraph and provide an instruction to the AI to rewrite the text in an arbitrary way (Figure 4); for example, 'rewrite this: *to be shorter*'. Participants in our formative study shared common difficulties with appropriate tone and style, while prior work shows that people with dyslexia will often rely on a thesaurus [47] or templates [13, 46] to achieve desired phrasing. Through custom instructions, users with dyslexia can specify their intentions for a passage and call upon the AI to select wording that meets that intention. Rewritten passages from the LLM were returned as several choices in an effort to maintain users' autonomy over the final passage.

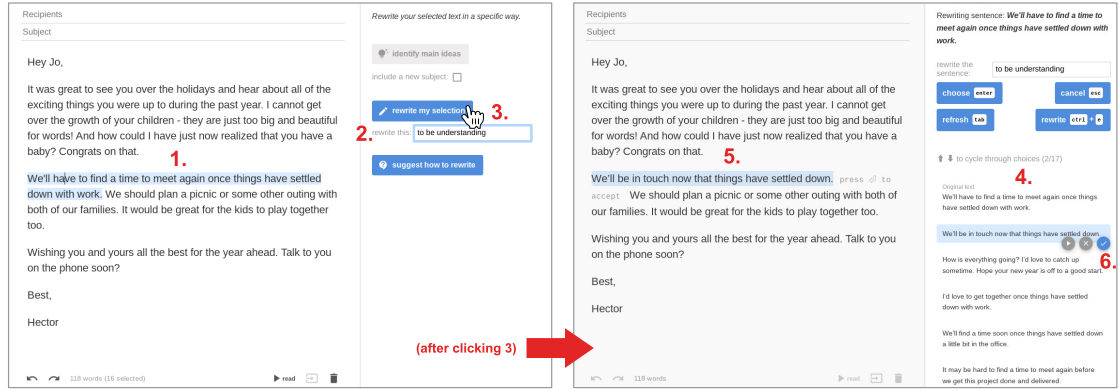

Fig. 4. **The Rewrite My Selection feature.** A user can select a piece of text (1), provide a custom instruction for changing it (2), and click *'rewrite my selection'* (3). Several rewritten choices from the AI will populate in the right-hand panel (4). Highlighting a choice with show a preview in the editor (5). For each choice, users can have it 'read aloud', discard it, or apply it to replace the original passage (6).

To implement this feature, we adapted the *related example prompting* method from [69]: rather than anticipating every possible user instruction and including these in the few-shot learning prompt, our prompt only contained a collection of examples *related to* instructions that we anticipated would be relevant to users with dyslexia for emails. Related examples are generally able to steer the model to completing an unseen, user-generated task [69]. Our examples included instructions for conciseness (*'to be simpler'*), tone (*'to be more polite'*), audience (*'to be more formal'*), and precision (*'to be more clear'*). Each example also contained the user's selected text paired with some text before and after the passage to provide context, and an ideal way to rewrite that passage according to the given instruction. When users press the *'rewrite my selection'* button, their current selection and surrounding text is appended to the end of the few-shot prompt and sent to the model; the model responds with several rewritten passages, shown to the user as choices.

### 3.3 Accessibility Considerations

Building a text-editing tool for users who find it difficult to parse and manipulate text presents an inherent design challenge. Although our primary goal for the LaMPost system was to demonstrate the functionality of LLMs for writing to users with dyslexia, we recognized that usability issues may impact their ability to fully evaluate the system. To mitigate this, we made several design choices to maximize usability for this population, and tested iterations of our design among members of our team who identify as having dyslexia. We used a sans serif font throughout the system because these have been shown to be the most readable and preferred by users with dyslexia [50]. To further improve readability, we incorporated sizing recommendations from prior work [54] suggesting a large font (18 points or more) with line spacing near the default value (1.0 units or 120% of the font size); based on feedback from our team, we chose an 18pt font size with 140% line spacing for the main editor panel. To support visual referencing, we paired most buttons with icons and added highlighting to the sentence surrounding the insertion point cursor (visible in Figure 4-1). Finally, for users that felt more comfortable listening to on-screen text than parsing it visually, we used the Web Speech API [2] to include a "read aloud" feature for the email's body, the generated outline, and each choice returned by the LLM.

## 4 LAMPOST EVALUATION

We evaluated the LaMPost prototype in a hands-on demonstration and practical email writing exercise. Our primary goals were to explore the potential ways that LLMs can be incorporated into the email-writing process of writers with dyslexia, and to assess users' perceptions of each of LaMPost's writing support features. In addition, we had secondary goals of understanding users' feelings of satisfaction, self-expression, self-efficacy, autonomy, and control while writing with LLMs, and to assess how exposure to AI terminology may impact these feelings.

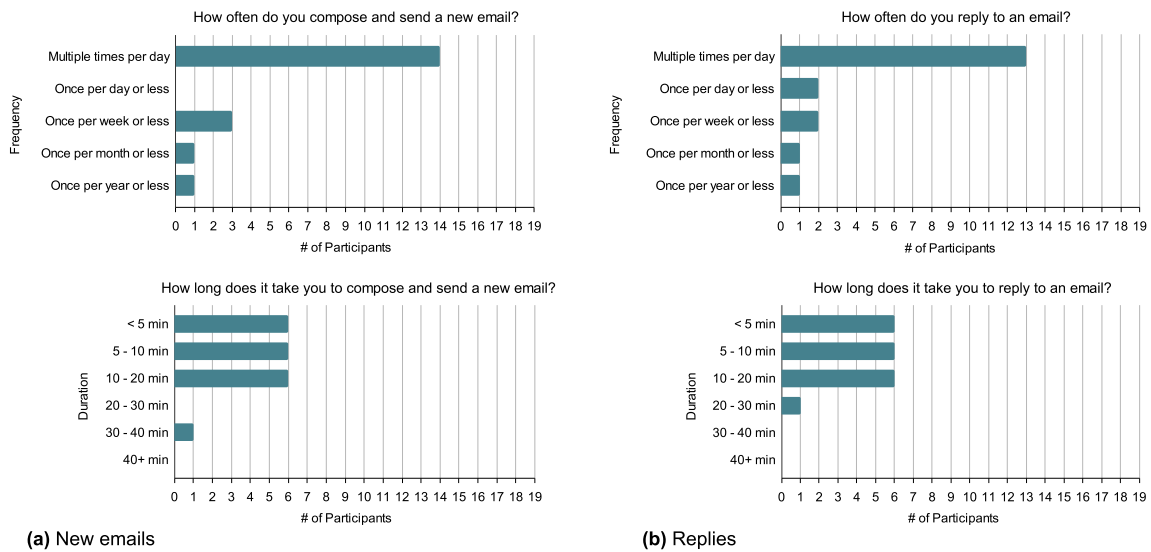

Fig. 5. Email-writing habits for the evaluation's 19 participants, showing frequency and duration when writing (a) new emails and (b) replies. Most participants reported writing new emails and replies multiple times per day and spending 20 minutes or less on each one.

### 4.1 Method

*4.1.1 Participants.* We recruited 32 participants via a survey shared with a large sampling pool maintained by our institution; 19 completed the study. All were based in the U.S. (*N*=16) and Canada (3); all said English was their preferred language to write. The recruiting survey asked if they had a dyslexia diagnosis, their emailing habits (Figure 5), and a series of demographic questions. We screened for experience writing emails (at least one per year) and self-reported challenges associated with dyslexia, but we did not require a dyslexia diagnosis to accommodate individuals without access to formal screening procedures. Fourteen participants reported having a formal dyslexia diagnosis and four reported discovering their dyslexia on their own; one participant did not specify. We aimed for balanced representation across gender and age categories, but we attained neither due to cancellations. Four participants identified as female, 14 as male, and one as non-binary. One participant was 18-24 years old, seven were 25-34 years old, and 11 were 35-54 years old. Participants were compensated with a $100 gift card for their time.

*4.1.2 Procedure Overview.* The evaluation procedure was split into three parts during a 75-minute period and conducted remotely due to the ongoing COVID-19 pandemic. First, we asked a few background questions to understand each participant's email-writing workflow, then provided a hands-on demonstration of the LaMPost system.[3] Second, we conducted an informal writing exercise in which participants freely used the LaMPost system to write at least one realistic email. Third, we asked semi-structured follow-up questions about the experience, then asked them to fill out rating scales evaluating the system's usefulness, consistency, and their own feelings while using the system—including satisfaction, self-expression, self-efficacy, autonomy, and control.

Maintaining feelings of autonomy during AI-assisted writing was a key design goal for LaMPost, following user concerns expressed during our formative study and in prior work [27]. However, as illustrated by emerging work showing the effect that metaphor choice can have on user perceptions [33, 36], our choices for LaMPost's presentation may influence the expectations, evaluations, and attitudes among users. Overt AI presence might be viewed as reducing autonomy, while obscured AI could resemble traditional computer-aided writing (*e.g.*, spelling/grammar check). To better understand how framing and the presence of AI metaphors can impact perceptions of an LLM-powered writing tool among adults with dyslexia, we segmented the system evaluation into two between-subjects conditions:

(1) **With AI metaphors** (*N*=9): We introduced LaMPost as an *"AI-powered"* email editor and used language throughout the session to present LaMPost's LLM as a personified AI agent providing writing assistance. For example, *"You can get a few suggestions from the AI for how it thinks you should write this differently"*, *"Hang on, the AI is thinking..."*, and *"This is feedback from the AI on how to improve your selection."*

(2) **Without AI metaphors** (10): We introduced LaMPost as an *"enhanced"* email editor and used language throughout the session to obscure the presence of an AI/LLM. The examples above in this condition: *"You can get a few suggestions for how to write this differently"*, *"Hang on while the system loads..."*, and *"This is feedback on how to improve your selection."*

The interface did not reference the system's underlying LLM mechanism, and we left it unchanged for both conditions.

Before the session, we asked participants to prepare two ideas for emails that they intended to write and felt comfortable sharing with us. To ensure email ideas were of sufficient length to conduct a substantive test of the system, we included an example: *"family newsletter on key events from 2021"*. We provided assurance that the writing exercises were only meant to provide a realistic experience of using the system, and that their performance would not be evaluated during the session. With participants' permission, all evaluation sessions were recorded for later viewing and analysis.

*4.1.3 Part 1: Background and Demo (25 min).* The first part of the study was used to learn more about each participant's current approach to email writing through a small set of interview questions and rating scales, and to demonstrate the LaMPost system's functionality. To begin the interview, we asked participants why they write emails, to share successes and challenges experienced when emailing, to recall a past instance of confusion among an email recipient, and to walk us through their email-writing process. Next, we asked participants to rate their confidence in their emailing ability (relates to self-efficacy [57]), ability to express themselves and their ideas (self-expression; *e.g.*, [38]), and their overall satisfaction with their emails. The three ratings were predicated by a positive statement (*e.g.*, *"I am confident in*

---

[3] Two participants (P12, P18) were unable to access the system due to an unknown technical issue. These participants dictated the content of the email and directions for using each feature to the researcher via a shared screen.

*the emails that I write.*") and followed by a 7-point scale from *"Completely disagree"* to *"Completely agree"*. For privacy, we shared the scales with participants via a linked form and offered to read each statement aloud if they desired: two participants opted for rating statements read aloud throughout the study; one read the statement aloud themselves; 16 read silently.

After participants had completed the rating scales, the researcher introduced the functionality of the system according to their assigned AI-metaphor condition in a hands-on demonstration. Participants opened the LaMPost system in a new tab and shared their screen with the researcher, who began by walking them through each element of the main editor panel: input for the email's recipient and subject, space for the body text, and buttons for "undo", "redo", "read aloud", and "clear". To make sure participants were following along, we asked participants to move their mouse to each element before it was introduced. To aid in demonstrating the LLM features, we included an additional "insert sample text" button that added a two-paragraph sample email into the body of the editor.

Next, we introduced each of the three LLM features (*Identify Main Ideas*, *Rewrite My Selection*, *Suggest Possible Changes*): we explained the feature's intended function, asked participants to try using it on the sample email, and explained any follow-up functionality associated with that feature. For the *Main Ideas* feature, this functionality included hovering over an idea in the structure to highlight its associated paragraph, clicking on an idea to hear it read aloud, and using the checkbox to generate a new subject line. For the *Rewrite* feature, this included selecting different options to see previews in the editor, hearing an option read aloud, deleting undesired options, and choosing an option to replace the selected passage. We gave a similar explanation for the options generated by the *Suggest* feature, with an additional note that a chosen suggestion could be sent as an instruction for the *Rewrite* feature if desired. After introducing each feature, we asked participants to share their immediate thoughts about it, any concerns with it, and if they thought they might use it when writing emails.

*4.1.4 Part 2: Writing Exercises (25 min).* After participants had been introduced to LaMPost, they used the system to write emails based on the ideas that we had requested for the session. To minimize stress associated with time constraints among writers with dyslexia [11, 30, 46] and challenges associated with learning an unfamiliar system, we did not require participants complete their email during the 25 minutes allotted. Instead, we told participants to write as much as they were able in the time provided and to freely use the LLM features as desired; if they finished with one email, they could try writing another.

We asked participants to "think aloud" throughout the writing exercise, and to freely voice any questions, observations, suggestions, or concerns about the system. If their actions were unclear for any reason, the researcher prompted them for an explanation. The researcher also provided limited answers to questions about the system's functionality with the appropriate language for each participant's AI-framing condition. To ensure that the participants had experience with each LLM feature (*Main Ideas*, *Rewrite*, *Suggest*), the researcher allowed them to write and use the system for at least five minutes before prompting them to try an unused feature, repeating until all features had been used at least once. We logged participants' complete use of the system, including: all typed additions, changes, and deletions; all buttons and LLM features used; all responses from the LLM; and all accepted LLM responses added to the document.

*4.1.5 Part 3: Follow-up Interview and Rating Scales (25 min).* Following the email writing exercises, we discussed the experience via semi-structured interview questions and rating scales. We began with questions to learn about participants' overall experience, asking them to compare their use of the system to their typical experience writing emails, and to share anything they found easier or more difficult than usual. Next, we asked for their opinions about the LaMPost system, including what they liked most about it, what needed improvement, and if they had any ideas for additional features that could assist them with writing. We used post-use rating scales to assess their overall impressions of the system and measure the impact of the AI-framing manipulation. The ratings targeted several concepts, including: usefulness and consistency of each LLM feature and the system overall; satisfaction with the system and the emails produced with it; and personal feelings of self-efficacy [57], self-expression (*e.g.*, [38]), autonomy, and control (*e.g.*, [22, 62]) while using the system. Each rating was predicated by a positive statement and followed by a 7-point scale from *"Completely disagree"* to *"Completely agree"*. After they had privately filled in the scales related to each concept, we asked a follow-up question related to the concept to capture the reason for each rating.

*4.1.6 Analysis.* We analyzed our qualitative data following the thematic coding process of Braun and Clarke [6] using a combined inductive and deductive approach. Prior to the study, we produced a set of deductive codes to categorize: existing email-writing practices; positives, negatives, and desired changes for each feature and the overall system; and feelings of self-efficacy, self-expression, autonomy, and control during and after system use. During data collection, three researchers produced session notes and observations and generated a set of inductive codes through analytic connection across participants. We used both sets to produce a final codebook containing a 3-level hierarchy: level-1 included high-level codes for writing practices, the overall system, and each feature; level-2 included positives, negatives, concerns for each; and level-3 included low-level codes based on our expectations following the formative study (deductive) and unexpected themes emerging from system use (inductive). One researcher used the final codebook to independently code transcripts for each of the 19 sessions, and resulting themes were organized into subsections and constructed to form our narrative. Our final codebook is provided as Supplementary Material. For quantitative data, we used pre-use ratings to characterize participants' existing feelings about writing emails. To compare how our between-subjects manipulation of AI framing shaped participants' perception of the system, we used a Mann-Whitney U test (two-tailed) to test for significance in post-use ratings for usefulness, consistency, satisfaction, self-expression, self-efficacy, autonomy, and control.

## 4.2 Findings

The core goal of our study was to understand the potential ways that LLMs can be incorporated into the email writing process of writers with dyslexia. In the following sections, we describe their current experience writing emails, reactions to our system and each feature, and the (lack of) an effect with the AI-metaphors condition.

*4.2.1 Current Email-Writing Experience.* We briefly explored participants' thoughts on writing emails to understand opportunities and challenges specific to this genre of writing. Figure 5 shows participants' email-writing habits; most participants said they wrote both new emails and replies daily (P5: *"In my busy season, up to 100 a day"*), though a few wrote less often (P19: *"once or twice a month"*). Participants said they primarily wrote emails for work communication; some also wrote personal emails to connect with family and friends or conduct service inquiries (*e.g.*, P8: *"doctor's appointments"*). In rating their level of agreement towards statements about writing emails (Figure 6), participants generally felt a strong sense of self-efficacy when emailing: the majority of participants (*N*=15) felt confident in their ability to write emails (*avg.*=5.05, *SD*=1.22). Participants also generally felt satisfied with their emails (*avg.*=4.89, *SD*=1.10) and that they could express themselves when emailing (*avg.*=4.79, *SD*=1.36), although the overall agreement towards each of these statements was more mixed.

When discussing their experiences writing emails, participants described similar challenges and mitigation strategies as we had heard about in our formative work and focus group, as well as those discussed in prior work (*e.g.*, [13, 42, 47]). Over half of participants (*N*=10) said they liked to draft emails outside of an emailing platform (*e.g.*, Microsoft Word, pen and paper) due to *"habit"* (P7), *"personal preference"* (P6), or *"to separate it from the anxiety of having to respond"* (P2). Within their preferred platform, participants described a common frustration of trying to convert information from their mind into writing—or, *"Putting what I'm trying to say in my brain into words"* (P13). Verbal communication was said to be easier than writing, and six participants relied on speech-to-text tools to dictate their thoughts; others had abandoned it due to errors: *"Softwares that I've used have had a slightly higher [error] percentage than my own [typed] inaccuracies"* (P5). For typing, nine participants mentioned predictive text, such as GMail's Smart Compose [35], as being particularly helpful to find their desired wording: *"I like the feature when it finishes what you are thinking; when you don't have to type it all out"* (P7). Common typed approaches included the "word faucet" strategy (*N*=5) identified during our formative study—*"I get all of my thoughts out so [...] I've got this blob of text"* (P10)—or bullet points (3) to map out high-level details. The remaining participants described unique initial drafting processes; for example, P6 described a linear, spontaneous method: *"I wing it. [...] I start the first sentence, then it's like, 'Okay, now I know what the second sentence will be.'"* In contrast, P8 preferred to write in a non-linear fashion: *"I start with just drafting the middle paragraph because I know that's the one that has the most information."*

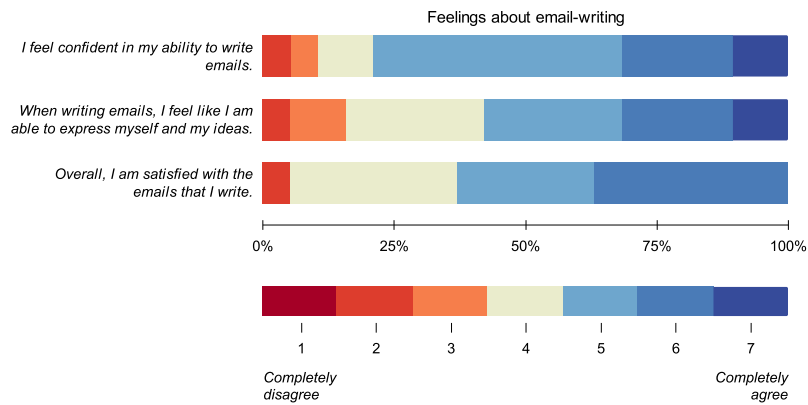

Fig. 6. Results of rating scales for feelings of self-efficacy, self-expression, and satisfaction during participants' existing email-writing process. Generally, most participants felt confident about email-writing, but had slightly more mixed feelings about self-expression and satisfaction.

After participants had moved their ideas to writing, additional challenges emerged during revising and proofreading. Spelling and grammar was the most common issue, and ten participants said they relied on a trusted spell checker, such as Grammarly, over their email platform to catch "real word" errors (P10: *"'there' and 'their'"*) and other mistakes. Cutting overly verbose, or *"wordy"* (P16), drafts to a succinct email was another frequent struggle (*N*=7). Participants recalled instances when they had used language that was misinterpreted by the email's recipient, most often due to a lack of clarity (*N*=7). For example, P13 recalled a recipient *"calling me on the phone that evening [because] he had no idea of what I was asking."* Tone was another common source of misunderstanding (*N*=6): *"They thought that I was just writing to them all mad and pissed off, when in reality, I was just explaining myself"* (P8). To check the email before sending, some participants (4) said they asked someone else to read the email, while others (3) said they listened to the message via text-to-speech. Several participants (7) mentioned they struggled to find enough time to adequately revise and proofread, especially when responding to urgent emails.

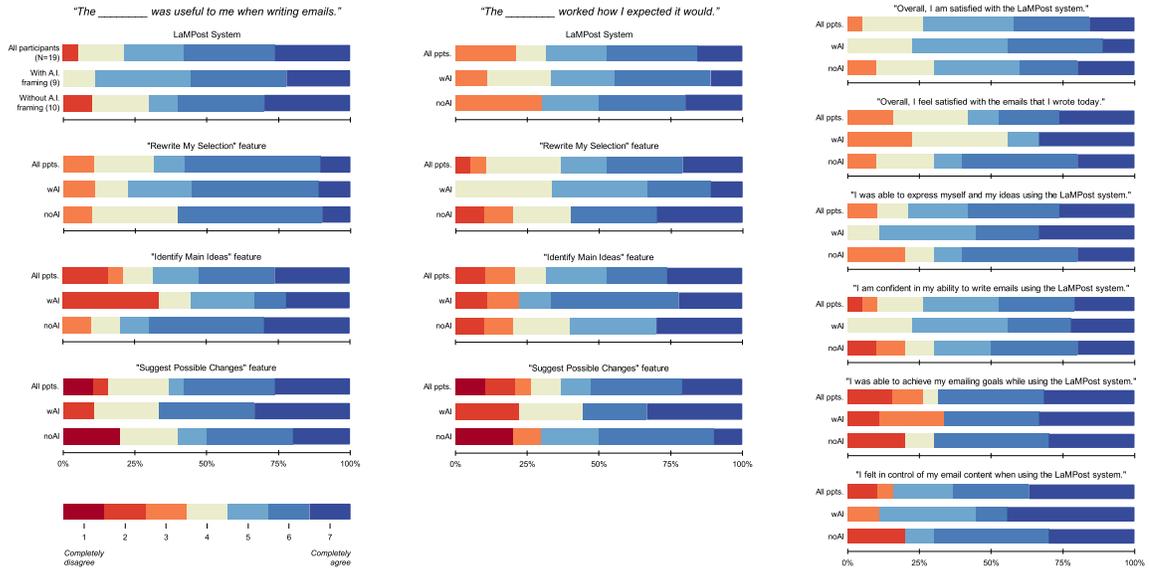

Fig. 7. Post-use rating scales among all participants, and across each of the AI framing conditions. We measured the usefulness (left) and consistency (center) of the system and each feature, and several of their feelings about using the system (right). In general, most found the LaMPost system useful while writing emails, appreciating the "Rewrite my selection" feature most of all.

### 4.2.2 Reactions to the LaMPost Prototype.

In the following section, we describe participants' responses to LaMPost's three LLM-powered features and the overall system based on usefulness ratings and feedback provided throughout the evaluation.

We asked participants to rate their level of agreement towards the usefulness of LaMPost's three LLM-powered features and the overall system. To begin, we used a two-tailed Mann-Whitney U test to compare usefulness ratings between the study's two AI-framing conditions—*i.e.*, *with* ($N$=9) and *without* (10) AI-related metaphors—but we did not find a significant difference for any of the ratings ($p$>0.05). Figure 7 shows results as a whole as well as results for each framing condition; we return to implications of the AI-framing experiment in the Discussion (Section 5.4).

Of LaMPost's three features, the **Rewrite My Selection** feature was rated highest for usefulness on average (*avg.*=5.26, *SD*=1.26); 13 participants agreed that it was useful for writing emails[4] and nine selected it as LaMPost's most useful feature. Participants said the primary benefit of the *Rewrite* feature was its capability to find satisfying and appropriate wording for an intended idea: *"You're able to get a start on what you're going to say and you can tweak your writing from there"* (P18). P6 liked the feature because the synonyms and alternative wording helped them to understand the meaning of their own writing—a similar benefit has been shown in prior work [49, 51]. While testing the *Rewrite* feature, participants saw further value for mitigating problems with language precision ($N$=7) and tone (4). For example, P1 said the feature would help to make his ideas more compact—*"I end up typing like three or four sentences to explain something simply"*—while P13 liked that he could *"sculpt the email to the audience"* via instructions to rewrite the text to be more *"business-like"* or *"laid-back"*. However, several participants took issue with the feature's implementation and functionality: the most common concerns were inaccuracy and noise from the model ($N$=13) and its overly numerous choices (11); four participants selected *Rewrite* as LaMPost's least useful feature as a result. We outline these and other concerns in greater detail in the following sections.

---
[4] Ratings $\geq$ 4 on 7-point agreement scale.

The **Identify Main Ideas** feature was rated the second highest on average in terms of its usefulness (*avg.*=5.11, *SD*=1.82); 13 participants agreed that it was useful to some extent, but only four participants selected it as the most useful of LaMPost's three core features. These participants identified the primary value of the feature's visual outline as validating that their writing contained the intended meaning for the email. For example, P4 said the feature was the system's *"biggest selling point"* because it would allow *"the ability to see, and make an independent verification, that I'm hitting the points that I want to hit."* While the feature's automatic summarization was sufficient to capture the *"gist"* (P4, P16) of each paragraph, four participants desired key details added to the outline: *"It definitely helped me recognize what I was talking about, other than the fact that it missed time-sensitive information"* (P2). Some participants (*N*=8) also voiced concerns over the summarizations feeling *"sterile"* (P17), or missing the emotion contained in the text. For example, P9 felt dissatisfied when a three-sentence paragraph—written to thank his project collaborators and share a draft of their production—was reduced to an outline item that stated, *"We have a video."* Nine participants selected the *Main Ideas* feature as the least useful overall, including four who said that they could not see a practical use for it in their writing process: *"It seems very obvious"* (P6).

Notably, the option to **generate a subject line** received a very positive response. Because LaMPost created the subject line from the *Main Ideas* outline returned by the LLM, we chose to pair the subject as an optional checkbox attached to the feature and did not ask participants to rate its usefulness separately. However, several participants—who otherwise felt tepid about the visual *Main Ideas*—saw value in a separate automatic subject feature: *"I always leave the subject line blank. [...] Nothing fits what I would be thinking"* (P19). Most participants thought LaMPost provided accurate subject lines for their emails, but a few (*N*=3) found issues with the framing of the subject. For example, P5 wondered why LaMPost added the subject line *"Invitation to [Name]"* for his email informing the individual that they would not be receiving an invitation: *"It makes sense, but given the context, that's not quite right. The idea I'm trying to get across is quite the opposite: something about bad news, or 'Information about Upcoming Event'."*

Reactions to the **Suggest Possible Changes** feature were mixed in terms of its usefulness (*avg.*=5.05, *SD*=2.03); six participants said it was the most useful feature, and six said it was the least useful feature. Twelve participants at least partially agreed that the feature was useful, and they identified the feature's primary benefit as support for fixing a detected, but unclear issue—*i.e.*, *"Times where I'm like, 'Something doesn't sound right'"* (P15). For example, in an email to his wife, P19 used the *Suggest* feature on a sentence that *"didn't feel natural"*, yielding a positive result: *"Here we go, 'Rewrite this phrase to be more romantic'. That's kind of what I was getting at."* A few participants further identified the *Suggest* feature's value in tandem with the *Rewrite* feature to provide possible boundaries for the latter's custom instructions: *"It helped narrow down the options that [Rewrite] could do"* (P12). However, not all participants agreed that the *Suggest* feature was helpful. Two participants were unsatisfied with feature's limited scope, saying it, *"Raises more questions than answers"* (P16), and desired further explanations of the suggested changes: *"It's saying, 'Rewrite this sentence to be less wordy'. Is it telling me that the sentence is wordy? And why is it at the top of the list? [...] And is it really wordy? There's two sentences in the paragraph."* (P10). Other frequent issues identified for the *Suggest* feature were similar to those for the *Rewrite* feature: several participants (*N*=9) were concerned about inaccuracy and noise in the results, while others (5) mentioned the *"overwhelming"* (P19) quantity of choices. We describe these issues further in the following sections.

Finally, while responses to individual features were varied, most participants identified at least one feature that was useful to them. As a result, the usefulness of **the LaMPost system overall** was rated fairly high following the

email-writing exercise (*avg.*=5.53, *SD*=1.38). P13 summed up the system's utility as, *"It allowed me to validate for myself that the point I'm trying to make is actually getting across, and it gives me the opportunity to rewrite it if it's not."* In general, participants were fond of LaMPost's capabilities for automatic, content-specific support (P19: *"like it's an extra person helping you"*) and being able to direct that support to the scale of their choosing: *"I could do sentence-by-sentence, or paragraph-by-paragraph"* (P10). However, participants were split on how the system's current capabilities would help during day-to-day emailing. For example, P14 explained the contexts where he saw it having value: *"Where I'm having an informal dialogue with a couple teammates—I don't need to spend this amount of time rewriting three sentences. [...] [But] I can see the benefit of using something like this when writing emails to a VIP, or if I'm trying to convey a complex topic."* In total, five participants mentioned LaMPost as potentially increasing the time required to write emails, but eight participants saw an overall time-saving benefit: *"I don't have to rewrite it, and rewrite it, and rewrite it. [...] I get five minutes back of my life"* (P4). Notably, one participant disagreed that the overall system was useful for his needs: *"It's more about potentiality versus reality. If you mean in its current state, I would say not very useful. If you mean if I see how it could be extremely useful, then it's on the complete opposite end of the spectrum"* (P5).

In the following sections, we outline participants' concerns with the system's current state in greater detail—highlighting their concerns over accuracy and choices in particular—and discuss further improvements they requested to better assist with their email-writing process.

*4.2.3 Concerns Over Accuracy and Noise.* One of the most common issues highlighted by participants throughout the evaluation was unhelpful, inaccurate, or *"nonsensical"* (P8, P11) results they received from the model. Anticipating the potential for instability in our few-shot learning prompts when given unseen tasks [71], we included an additional *consistency* rating for each feature and the overall system.[5] A two-tailed Mann-Whitney U test to compare each consistency rating between the evaluation's AI-framing conditions did not yield a significant difference (*p*>0.05) for any rating; Figure 7 shows results for each consistency rating as a whole, as well as results for each framing condition.

Poor results were most apparent from the *Rewrite* and *Suggest* features, where our few-shot prompts tasked the LLM with providing new suggestions for rewriting or changing the text; generative tasks that are more unpredictable than the relatively constrained *Main Ideas* summarization. This unpredictability sometimes led to striking "hallucinations"—factually incorrect or non-existent content generated by the LLM [20, 48]. For the *Suggest* feature, the hallucinations amounted to suggested changes that were irrelevant for the selected passage, such as *"Rewrite this sentence to not use so many commas"* given to P11: *"There's not a comma in it."* Some participants were able to tolerate a few irrelevant suggestions (P5: *"It sparks other ideas for how to rewrite things"*), but others could not: *"If this is consistent, I'm going to think, 'Why am I bothering to use [this feature] in the first place?'"* (P10)

In contrast, hallucinations within the *Rewrite* feature amounted to seemingly relevant yet imaginary details added to the rewritten passage. For example, when evaluating a choice containing the phrase, *"Maybe that nice patio you were telling me about,"* P12 noted, *"The original text doesn't mention a patio."* Although he removed the option from his remaining choices, he wondered where the system had found this information, why it had chosen to include it in the email, and whether or not he would be able to catch more *"wrong info"* in the future. In one instance, the hallucinated details carried a deeper personal implication: P6 described feeling unsettled by an option containing the phrase, *"I've also posted on local Facebook,"* for an email informing a friend about schools in the area: *"That's kind of one of those 'gaslight'-y inaccuracies that unnerve me because I would never be on Facebook. Although now, I'm like, 'Wow, wait a*

---

[5]A 7-point rating scale measuring level of agreement toward *"The [feature / system] worked how I expected it would."*

*minute. Duh. Of course that would be a place to find out about this stuff.' [...] It's almost like a suggestion or an assistant telling me to go do that. That's not necessary."*

Hallucinations weren't the only source of inaccuracy for the *Rewrite* feature. Four participants commented on choices that did not satisfy the instruction that they had given; for example, after the instruction *"Rewrite this text: to sound more detailed"* gave a few concise results, P2 noted, *"To me, this [instruction] implies more text than I wrote."* Although the system removed repeated suggestions in the final choices displayed to each user, three participants mentioned seeing overly similar results: *"This one is just the same sentence that I've typed"* (P3). Two participants noticed that some rewritten choices had removed important details contained in the original text. Although many participants were ultimately able to use the *Rewrite* and *Suggest* features to find their desired wording or query for helpful changes, the process of sifting through inaccurate or irrelevant results was both *"time wasteful"* (P7) and cognitively demanding (P14: *"I need to put in a lot of focus"*).

*4.2.4   "The Paradox of Choice".*   The LaMPost system included two extremes with regard to choice: the *Rewrite* and *Suggest* features displayed numerous choices to users (*i.e.,* 15 responses returned from the model, minus duplicates), while the *Main Ideas* and subject line returned the model's first response each time. In this section, we discuss participants' concerns with each extreme, and their suggestions for improvement.

Twelve participants across both the *Rewrite* and *Suggest* features voiced concerns over the sheer volume of choices to parse. Two participants described this as *"the paradox of choice"* (P5, P6), referring to the psychological concept that large sets of choices can feel overwhelming and lead to poor or unsatisfying final selections [43, 56]. *"My curiosity would just lead me into spiraling and reading hundreds of these possible choices. And maybe there'd be more self-doubt from that spiraling"* (P6). Yet others found numerous options were helpful—even necessary—to find a correct option amidst undesirable or inaccurate results from the model. *"One that was suggested was pretty much on point. The rest of them are either a little off, or would require a bit of rewriting"* (P4). Still, most participants agreed that the number of options displayed at a time should be considerably reduced, suggesting around *"three to four options"* (P19), or *"four to five; nothing I need to scroll through"* (P2). One promising idea for filtering through a reduced number of options came from P6: *"The choices should be sorted by word count, just five or six choices, tops. And if I 'X' out one of those, it gives me a new one."*

Additional tensions around user choice emerged from the open-ended possibilities for the *Rewrite* feature's free-form instructions. Six participants spoke of the immense value in being able to write their own instructions, such as P10: *"I don't tend to think in the way that commonly is spoke, so being able to put down my thoughts and have an AI [respond]— [...] It's like having a thesaurus for someone's thought."* Others voiced concerns about this capability, including doubts towards complex inputs (*N*=3), its handling of misspellings (P8, P13), and potential dangers with misguided instructions: *"If it's two in the morning and I'm emailing back and forth with support, I might want to type something like, 'Make it sound mad'"* (P4). These concerns led to six participants requesting pre-written instructions of general changes included alongside the open-ended input—and distinct from the context-specific instructions that could be queried from the *Suggest* feature. For example, P13 said it seemed tedious to type certain instructions: *"I had to go to my spell checker on my phone to figure out how to spell the word 'formal' before I could even use it. [I want] a drop-down of some of the most basic ones: [...] 'make it shorter', 'make it longer', 'more professional', 'more casual'."*

The other choice extreme—the single result of the *Main Ideas* and subject line feature—also raised concerns, although less frequently. While pointing out that key details were missing from the visual *Main Ideas* outline, two participants wondered why they were unable to choose the main ideas themselves. For example, P1 criticized the feature for *"missing* *the forest for the trees"*—concentrating on one thing while ignoring the rest—and offered a solution: *"An option where I could highlight a section and then tell it, 'This is what I want you to work your identification around."'* Three participants questioned why the generated subject line was returned as single result, such as P8 after several repeated attempts: *"It would be helpful to see a list of options."*

*4.2.5 Feelings About Email-Writing with LLMs.* The secondary goal of our evaluation was to understand how writing with LLMs can impact personal feelings of satisfaction, self-expression, autonomy, and control among writers with dyslexia. We asked participants to rate their level of agreement towards these feelings within the context of the LaMPost system (Figure 7); a two-tailed Mann-Whitney U test to compare each consistency rating between the evaluation's AI-framing conditions did not yield a significant difference ($p$>0.05) for any rating. In the following section, we briefly discuss each feeling based on the results of the rating scales and relevant comments provided throughout the study.

Participants generally felt satisfied with the system (*avg.*=5.26, *SD*=1.14), but they voiced several concerns over the LLM's accuracy and our implementation of choices (described in the sections above). Most participants also felt satisfied with what they wrote using the system (*avg.*=5.16, *SD*=1.52), but this rating varied depending on the complexity of their chosen emailing task: the majority (*N*=12) were able to finish a full email within the 25-minute writing period (P8: *"I would actually send this out to one of my friends"*), while the rest (7) were unable to complete their task due to limited time. This variation was also apparent in their ratings of personal autonomy (*avg.*=5.26, *SD*=1.88), where participants had ranging levels of agreement about whether or not they had achieved their emailing goals. Autonomy ratings generally aligned with email completion; *e.g.*, P11: *"I got my point across"* vs. P6: *"I still had a few more ideas to express."* However, two participants that did not finish writing their emails said the system still helped them to achieve other personal goals: *"staying on task"* (P16) and *"keeping focus"* (P2).

Participants' sense of self-efficacy using LaMPost was fairly high (*avg.*=5.26, *SD*=1.41), and 14 participants at least partially felt confident writing emails with the system: *"It would be like a having a proof-reader along with me"* (P3). Those that felt less confident expressed doubts towards the system's accuracy: *"I think the AI would start breaking down if it really had to compute more and more"* (P1). Participants generally rated their sense of self-expression with LaMPost high (*avg.*=5.53, *SD*=1.34), particularly due to the suggestions provided by the *Rewrite* feature: *"I was able to look and say, 'Well, this is more of how I would speak,' or, 'This is more of how I would want the email to sound'"* (P18). However, four participants did not find LaMPost helpful for expressing their ideas, such as P8, who gave a neutral rating: *"The AI is limited on what it receives from the user, and it cannot explain much of the email if I don't give it the [information]. [...] And I don't always know what to write."*

AI-assisted writing systems can introduce complex questions around whether the user or agent is ultimately in control over the produced work [27]. Participants largely felt in control over their email content while using LaMPost (*avg.*=5.58, *SD*=1.56), and positive responses were closely associated with the ability to filter through the model's results and make final decisions over changes to the text. For example, P15 felt a strong command over his work: *"Even though the system gave suggestions, in the end, I'm the one that's deciding what it's going to say."* However, three participants disagreed with the statement, citing limited control over the generated subject line (P5), the opaque source of the writing suggestions (P17), and the inability to troubleshoot after undesirable results: *"It wasn't acting on it's own or anything, [...] but there was no actual, refined control over each process once I put it into motion"* (P1).

While not included in our rating scales, our evaluation also elicited feelings around privacy and trust. For privacy, seven participants voiced concerns about the system reading and storing their personal data: *"What is it doing with the information that I put in there? Where does that go?"* (P17) To protect their sensitive information, P9 desired *"the option* *to not have it read it"* while P4 requested a clear explanation of the system's data storage policy. With regard to trust, a few participants highlighted the importance of building their trust in the system, especially if they were relying on it for support in challenge areas. P13 used his own difficulty with determining writing's tone to demonstrate the issue: *"If you wrote the same statement three different ways—one professional, one casual and one romantic— [...] I would read them and they'd all look exactly the same to me. [...] So if [the system is] saying, 'I can take this sentence and make it more businesslike', I'm going to accept whatever you're offering me."*

*4.2.6 Additional Features and System Improvements.* Participants suggested several changes for the LaMPost system. In addition to overall improvements to the system's accuracy, six participants thought the system should include personalization–either *"learning through prior conversations"* (P6) or *"narrow[ing] down what words a user would use"* (P19) by learning from choices over time. Some element of personalization was needed to capture the writer's voice, according to P17, *"Rather than it being a canned response from the computer."* For the LaMPost interface, the most common requests related to how each of the LLM-based features presented results to the user. Participants desired fewer choices for the *Rewrite* and *Suggest* features (*N*=12), and more choices for the *Main Ideas* feature (3). Five participants requested more features to track their progress and the system's results during each writing session, including a *"changelog"* (P6), the ability to save and favorite individual choices (P18), and creating action items from the *Suggest* results (P2). Further requests for the *Suggest* feature included an *"explanation"* button for each result (P10, P16) and suggestions generated in real time (P4).

Drawing from their preferred approach of writing emails from bullet points, four participants wondered if the system could generate the body of an email from a given outline (*i.e.*, implementing the *Main Ideas* feature in reverse): *"I write the three main topics that I want and the system writes [an email] around them"* (P9). Three participants also thought the visual *Main Ideas* outline would be useful for reading other people's emails, such as, *"If I'm late to a meeting and they're referring to an email that I have not read"* (P8). Six participants appreciated the "Read aloud" feature (P12: *"I wish more things could just be clicked on and read back to me"*), but they requested options to change the speed (P14, P15) and improvements to make the *"mechanical"* (P6, P12) voice sound more natural. Finally, four participants requested improvements to spelling and grammar detection, preferring to use existing *"autocorrect"* implementations on mobile operating systems and native email clients.

*4.2.7 Summary.* All participants identified one of LaMPost's features as being potentially useful to them when writing emails, and the *Rewrite My Selection* feature was chosen as the most useful overall. However, accuracy and quality concerns limited the practical usefulness of many features, despite LaMPost using today's state of the art models and prompting techniques. Quality concerns (and a desire to support writer autonomy) led us to present generated text as many choice for users to select from; however, a top-N approach rather than a top-1 approach can be particularly challenging for end-users with dyslexia due to the additional reading and cognitive load challenges.

## 5 DISCUSSION

The LaMPost system provided a testbed to explore the feasibility and potential of LLMs as AI-powered writing support tools through three features to meet high-level writing needs: *Identify Main Ideas (+ subject line generation), Rewrite My Suggestion,* and *Suggest Possible Changes.* Our results indicate that LLMs with optimal output hold potential to assist with email-writing tasks, and our evaluation highlighted several promising routes for future explorations of AI-assisted writing—including the popularity of the controllable *Rewrite* and subject line features. However, our features as-is did not surpass participants' accuracy and quality thresholds, and as a result, we conclude that state-of-the-art LLMs (as of early 2022) are not yet ready to fulfill the real-world needs of writers with dyslexia. Surprisingly, we found no effect in the use (or non-use) of AI metaphors on perceptions of the system, nor on feelings of autonomy, expression, and self-efficacy when writing emails. As a whole, our findings yield insights on the benefits and drawbacks of using LLMs as writing support for adults with dyslexia. Below, we discuss implications of our findings and opportunities for future work.

## 5.1 Implications for Designing Dyslexia Support with LLMs

When evaluating LaMPost, we found that some users with dyslexia desire personalized writing support, where the system would learn the user's preferred diction and tone from past writing samples and apply it when producing rewrites and suggestions for their current work. A personalized system may enable writers with dyslexia to express themselves more naturally, and results filled with familiar vocabulary and phrasing may be easier to parse. However, since personalization requires access to the data of individual users, it may also run in conflict with their privacy choices [31]—a concern that was expressed by participants in our work. What degree of personalization is preferred, and how much writing data is needed to implement it? Should the system collect this data automatically (*e.g.*, from an email platform's "Sent" folder), or are users willing to identify and share specific writing examples themselves? Can complex privacy policies be made more accessible through built-in text-to-speech or writing with simple phrasing that meets visual preferences for reading? There are opportunities for future work to incorporate personalization in AI-powered writing tools to support the needs of writers with dyslexia, but researchers must consider the potential trade-offs between these systems and the privacy preferences of users.

Confirming and extending prior work on ownership in AI-assisted writing [27], we found that it was important for users with dyslexia to feel in control of AI-produced writing. To make the final decision over changes to the text, participants with dyslexia in our formative study requested writing suggestions be delivered as several options, and we incorporated these findings in the design of the LaMPost system's *Rewrite* and *Suggest* features (but overlooked choices for the subject line). Our evaluation reiterated the importance of user choice, but also highlighted usability issues with LaMPost's large list of options: making a choice—especially from similar and/or lengthy options—was time-consuming, cognitively demanding, and overwhelming to some users. Our findings suggest these issues can be reduced by displaying fewer options, adding more variation, sorting among displayed options (*e.g.*, by length), and allowing users to save and return to options later on. Although the *Rewrite* feature's open-ended instruction was identified as a valuable mechanism for expressing users' intention, it led to further usability issues: some users had difficulty finding the words to express their desired changes while others encountered spelling and grammatical errors. Additional scaffolding to overcome these issues while instilling a sense control could include pre-written instructions, dictation options for users that prefer speaking, or probing questions to help users narrow down their revising goals. Our work highlights possible usability pitfalls with control-enabling mechanisms, and future work should seek to balance each users' sense of control with their ability to fully leverage the system's features.

Assessing the results of an AI-enabled assistive technology can present obstacles for primary users when the data is inherently not accessible [25]; likewise, assessing the quality of text-based results from an AI-enabled writing system presents challenges for users with dyslexia. Additional assessment mechanisms may be needed to support this population: LLMs often produce factually incorrect "hallucinations" [20, 48], and have been shown to inherit biases present in their training data (*e.g.*, internet posts) [21]. We did not encounter offensive results in our work, but certain *"nonsensical"* results prompted LaMPost users to request explanations (*e.g.*, question mark buttons) to assist with determining the suitability of results for their writing. User feedback options to flag incorrect or unacceptable language may offer a promising solution for developers to make targeted fixes. Increasing the transparency of an automated system can foster trust among users and drive continued use [23], and further trust can be gained from performance improvements. However, our work suggests that with sufficient trust and confidence in the system, users with dyslexia may feel less inclined to invest their time and energy towards analyzing each result—increasing the risk for harm. To address this, future work should explore potential safeguarding methods; for example, before sending an email, a system could perform a final check of all machine-written text, and ask the user to verify that this text seems accurate.

## 5.2    New Datasets May Improve Support

Our LaMPost evaluation highlighted users' limited tolerance for inaccurate or unhelpful LLM results when writing emails. We chose to use a pre-trained model as a generalized base for the LaMPost prototype, and used few-shot prompting [16, 69] to demonstrate each task from several exemplars of the task performed on generic email text. Although our exemplars contained text pulled from real emails that covered a range of topics, these were completed emails that had been produced by writers of unknown lexical ability—examples that may not have reflected the characteristics of early drafts produced by participants in our study. Further work is needed to understand if state-of-the-art LLMs show the same limitations when employed as email-writing support tools among the general population, and whether or not LaMPost's outlining, rewriting, and suggesting features (or similar) can benefit this audience.

   If we had constructed prompts with a more representative baseline of participants' writing, the quality of LaMPost's results may have improved during our study. To our knowledge, however, a public corpus of writing samples produced by adults with dyslexia does not exist. To maximize the potential of AI-assisted writing tools for this demographic, future work could collect samples of writing from users with dyslexia. A small dataset could be added to few-shot learning prompts to improve results from pre-trained LLMs, while a sufficiently large corpus could be used to train smaller, specialized models for high performance on constrained tasks (*e.g.*, summarizing main ideas). An ideal dataset should include snapshots from different times during the writing process, and give consideration to the varied writing approaches described by participants in our study. As an example, an early email draft from a user that prefers to start writing by outlining key points is very different from one dictated via a speech-to-text tool, and each draft will develop differently over time—yet users desire a tool that can support their needs across all stages of writing. Constructing this dataset would be a complex task, but it may be achievable through usage logging during future prototype evaluations with a diverse population of participants.

## 5.3    Lessons Learned in Few-Shot Prompting for LLMs

Before settling on the core features in LaMPost, we experimented with other high-level writing features identified during formative research as having potential benefit to writers with dyslexia; we present them here to show the limits of few-shot prompting methods and potential design opportunities for future work.

   Following positive responses toward personified "digital writing companions" in our formative study, we explored how LLMs could be used for a collaborative, conversational writing experience. We developed a prototype for emails drafted through an instant messaging interface that could leverage the LaMDA model's dialog-based prompting format. After users supplied a short statement explaining the purpose of the email, the LLM-as-chatbot would generate probing follow-up questions to capture each detail to be included in the message; in a separate panel, users could watch as the LLM gradually constructed an email from each piece of relayed information. However, chatting with the LLM exposed its tendency to propose operations that it could not functionally perform (e.g., *"Do you want me to remove that* *information from the email?*"), and we could not determine a reliable method to limit these misleading responses. We faced further issues with transforming the question-and-answer chat stream into a meaningful email.

Inspired by an individual in our formative study that preferred to write by expanding an initial bulleted list of key points (a practice also described in our evaluation), we explored a feature that would automatically draft an email from a given outline. We created several exemplars containing an outline of 3-7 bulleted ideas, and a complete email containing the outline's ideas reordered and expanded into full sentences. While the exemplary email added transition language and some light rephrasing, it did not include any new information; despite this, the model tended to "hallucinate" additional information that did not exist in the original bulleted list (*e.g.*, making up names of individuals when none were given; adding unrelated information from the prompt itself). We attempted to constrain the language of each exemplary email to more closely match that of the bulleted ideas, but this reached an extreme when the expanded "emails" were word-for-word reproductions of the given outline; while it successfully constrained the LLM's imagination, we decided this functionality would be of little value to users. Future efforts could try adding empty spaces throughout the exemplary email in place of any new language: the LLM will likely copy this stylistic choice and users may feel comfortable filling in specific details themselves.

The third attempt drew from the "word faucet" approach described in our formative study by exploring automatic ordering and structuring for a long, disorderly text produced by, *e.g.*, speech-to-text dictation. We intended for the feature to improve a given "block" of text by arranging ideas into logical order and separating the writing into discrete paragraphs. Here, we ran into issues with the model's limited context window: because it accepts prompts with a finite length, we could include just one or two exemplars to demonstrate the structuring task—leading to inconsistent results. To overcome this, we tried breaking up the task via a prompt-chaining process [67]: first, summarizing the key ideas contained in the "block" (creating a paragraph structure); next, connecting each sentence with one of the key ideas (building each paragraph's content); and finally, arranging the paragraph's sentences into a logical order. We decided this feature was not practical for our user study: the chaining process required several minutes to complete, and deconstructing the input in this way stripped each sentence of meaningful surrounding context. However, as LLM context windows increase in size and chaining paradigms mature, this feature may be achievable in the future.

### 5.4 Limitations of Framing *with* and *without* AI Metaphors

Our evaluation explored whether or not the presence of AI metaphors impacted users' perception of the tool, but we did not find statistical significance for any rating. Based on prior work showing the effect of different conceptual metaphors on perceptions of automatic systems [33, 36] and concerns over autonomy in human-AI writing [27], we had hypothesized that a user's knowledge that LaMPost was AI-powered could reduce their sense of autonomy during use. A lack of a significant result for any of the subjective ratings in our AI metaphors conditions may be taken as positive outcome for public attitudes towards AI writing support tools; knowledge of the AI proved neutral for users' sense of ownership over the text, and for each of our other measured feelings. However, this result may have also been caused by systematic error in the form of small sample sizes for each condition ($N$=9 *vs.* 10) or study fatigue [58] (most rating scales were administered at the end of a 75-minute video conferencing session). We did not attempt to measure participants' prior experience with AI; per our institution's recruiting guidelines, participants were not affiliated with the technology industry and we did not expect them to have deep familiarity with AI before the study. Yet a few participants in the *without AI* group assumed the system was powered by AI and described it as such without our mentioning it; this could further indicate that our manipulation was not designed correctly (*e.g.*, AI was implied despite non-specific vocabulary), or it may simply reflect an increasing awareness among the public toward the capabilities of AI and the likelihood that AI is powering many new products and experiences—despite many lay users not understanding which specific technologies are AI-infused [34]. Future research should attempt a deeper exploration of the effects of presenting writing tools as AI-forward *vs.* obscured to better characterize the possible impacts of each on user attitudes—particularly end-users with dyslexia—and provide design guidelines for off-the-shelf systems that begin to incorporate this technology.

## 5.5 Limitations

We identified several issues as a result of conducting a remote lab evaluation amidst the ongoing COVID-19 pandemic. We recruited 32 participants, but only 19 completed the evaluation; this caused an unbalanced demographic representation in our data and a smaller-than-planned sample size for each condition of our between-subjects AI-framing experiment. The remote nature of the study also limited our ability to control the testing environment. An unknown technical issues prevented two participants from accessing the system and had to dictate the email's content to the researcher via screenshare. Other participants required support with setup and troubleshooting, which reduced their time using the system during the writing exercise, and added variation to the average duration spent writing LaMPost. While all participants tested each of LaMPost's features during the exercise, seven participants were unable to complete their writing in the time provided. Finally, participants were unable to experiment with LaMPost for different email topics and audiences (*e.g.*, work vs. personal), potentially skewing their responses according to the complexity their chosen email writing task.

## 6 CONCLUSION

In this paper, we introduced LaMPost, an email-writing interface that explored the potential for large language models to power writing support tools that address the varied needs of people with dyslexia. LaMPost introduced AI-assisted writing features inspired by the needs of adults with dyslexia, including *rewrite my selection*, *identify main ideas* (with subject line generation), and *suggest possible changes*. Additionally, we contributed insights from an evaluation of LaMPost with 19 adults with dyslexia. Our study identified many promising routes for further exploration—including the popularity of the "rewrite" and "subject line" features—but also found that state-of-the-art LLMs (as of early 2022) may not yet have sufficient accuracy and quality to meet the needs of writers with dyslexia. Surprisingly, we found no effect in the use (or non-use) of AI metaphors on users' perceptions of the system, nor on feelings of autonomy, expression, and self-efficacy when writing emails. Our findings yield further insight into the benefits and drawbacks of using LLMs as writing support for adults with dyslexia and provide a foundation to build upon in future research.

## REFERENCES

[1] 2017. Dyslexia FAQ - Yale Dyslexia. https://dyslexia.yale.edu/dyslexia/dyslexia-faq/.

[2] 2021. GitHub Copilot: Your AI pair programmer. https://copilot.github.com/.

[3] Damien Appert and Philippe Truillet. 2016. Impact of Word Presentation for Dyslexia. In *Proceedings of the 18th International ACM SIGACCESS Conference on Computers and Accessibility* (Reno, Nevada, USA) *(ASSETS '16)*. Association for Computing Machinery, New York, NY, USA, 265–266.

[4] Thomas Armstrong. 2011. *The Power of Neurodiversity: Unleashing the Advantages of Your Differently Wired Brain (published in Hardcover as Neurodiversity)*. Hachette Books.

[5] Emily M Bender, Timnit Gebru, Angelina McMillan-Major, and Shmargaret Shmitchell. 2021. On the Dangers of Stochastic Parrots: Can Language Models Be Too Big?. In *Proceedings of the 2021 ACM Conference on Fairness, Accountability, and Transparency* (Virtual Event, Canada) *(FAccT '21)*. Association for Computing Machinery, New York, NY, USA, 610–623.

[6] Virginia Braun and Victoria Clarke. 2006. Using thematic analysis in psychology. *Qualitative research in psychology* 3, 2 (2006), 77–101.

[7] Tom B Brown, Benjamin Mann, Nick Ryder, Melanie Subbiah, Jared Kaplan, Prafulla Dhariwal, Arvind Neelakantan, Pranav Shyam, Girish Sastry, Amanda Askell, Sandhini Agarwal, Ariel Herbert-Voss, Gretchen Krueger, Tom Henighan, Rewon Child, Aditya Ramesh, Daniel M Ziegler, Jeffrey

Wu, Clemens Winter, Christopher Hesse, Mark Chen, Eric Sigler, Mateusz Litwin, Scott Gray, Benjamin Chess, Jack Clark, Christopher Berner, Sam McCandlish, Alec Radford, Ilya Sutskever, and Dario Amodei. 2020. Language Models are Few-Shot Learners. (May 2020). arXiv:2005.14165 [cs.CL]

[8] Maggie Bruck. 1990. Word-recognition skills of adults with childhood diagnoses of dyslexia. *Dev. Psychol.* 26, 3 (May 1990), 439–454.

[9] Daniel Buschek, Martin Zürn, and Malin Eiband. 2021. The Impact of Multiple Parallel Phrase Suggestions on Email Input and Composition Behaviour of Native and Non-Native English Writers. (Jan. 2021). arXiv:2101.09157 [cs.HC]

[10] S Cai, S Venugopalan, K Tomanek, A Narayanan, M R Morris, and M Brenner. 2022. Context-Aware Abbreviation Expansion Using Large Language Models. In *Proceedings of NAACL*.

[11] H E Cameron. 2016. Beyond cognitive deficit: the everyday lived experience of dyslexic students at university. *Disabil. Soc.* 31, 2 (2016), 223–239.

[12] Nicholas Carlini, Florian Tramer, Eric Wallace, Matthew Jagielski, Ariel Herbert-Voss, Katherine Lee, Adam Roberts, Tom Brown, Dawn Song, Ulfar Erlingsson, Alina Oprea, and Colin Raffel. 2020. Extracting Training Data from Large Language Models. (Dec. 2020). arXiv:2012.07805 [cs.CR]

[13] Christine Carter and Edward Sellman. 2013. A view of dyslexia in context: implications for understanding differences in essay writing experience amongst higher education students identified as dyslexic. *Dyslexia* 19, 3 (Aug. 2013), 149–164.

[14] Mark Chen, Jerry Tworek, Heewoo Jun, Qiming Yuan, Henrique Ponde de Oliveira Pinto, Jared Kaplan, Harri Edwards, Yuri Burda, Nicholas Joseph, Greg Brockman, Alex Ray, Raul Puri, Gretchen Krueger, Michael Petrov, Heidy Khlaaf, Girish Sastry, Pamela Mishkin, Brooke Chan, Scott Gray, Nick Ryder, Mikhail Pavlov, Alethea Power, Lukasz Kaiser, Mohammad Bavarian, Clemens Winter, Philippe Tillet, Felipe Petroski Such, Dave Cummings, Matthias Plappert, Fotios Chantzis, Elizabeth Barnes, Ariel Herbert-Voss, William Hebgen Guss, Alex Nichol, Alex Paino, Nikolas Tezak, Jie Tang, Igor Babuschkin, Suchir Balaji, Shantanu Jain, William Saunders, Christopher Hesse, Andrew N Carr, Jan Leike, Josh Achiam, Vedant Misra, Evan Morikawa, Alec Radford, Matthew Knight, Miles Brundage, Mira Murati, Katie Mayer, Peter Welinder, Bob McGrew, Dario Amodei, Sam McCandlish, Ilya Sutskever, and Wojciech Zaremba. 2021. Evaluating Large Language Models Trained on Code. (July 2021). arXiv:2107.03374 [cs.LG]

[15] Elizabeth Clark, Tal August, Sofia Serrano, Nikita Haduong, Suchin Gururangan, and Noah A Smith. 2021. All That's 'Human' Is Not Gold: Evaluating Human Evaluation of Generated Text. In *Proceedings of the 59th Annual Meeting of the Association for Computational Linguistics and the 11th International Joint Conference on Natural Language Processing (Volume 1: Long Papers)*. Association for Computational Linguistics, Online, 7282–7296.

[16] Andy Coenen, Luke Davis, Daphne Ippolito, Emily Reif, and Ann Yuan. 2021. Wordcraft: a Human-AI Collaborative Editor for Story Writing. (July 2021). arXiv:2107.07430 [cs.CL]

[17] Craig Collinson and Claire Penketh. 2010. 'Sit in the corner and don't eat the crayons': postgraduates with dyslexia and the dominant 'lexic' discourse. *Disabil. Soc.* 25, 1 (Jan. 2010), 7–19.

[18] V Connelly, E J Sumner, and A Barnett. 2014. Dyslexia and writing: Poor spelling can interfere with good quality composition. *Brookes eJournal of Learning and Teaching* 6, 2 (Dec. 2014).

[19] Vagner Figueredo de Santana, Rosimeire de Oliveira, Leonelo Dell Anhol Almeida, and Maria Cecília Calani Baranauskas. 2012. Web accessibility and people with dyslexia: a survey on techniques and guidelines. In *Proceedings of the International Cross-Disciplinary Conference on Web Accessibility* (Lyon, France) *(W4A '12, Article 35)*. Association for Computing Machinery, New York, NY, USA, 1–9.

[20] Jacob Devlin, Ming-Wei Chang, Kenton Lee, and Kristina Toutanova. 2018. BERT: Pre-training of Deep Bidirectional Transformers for Language Understanding. (Oct. 2018). arXiv:1810.04805 [cs.CL]

[21] Jwala Dhamala, Tony Sun, Varun Kumar, Satyapriya Krishna, Yada Pruksachatkun, Kai-Wei Chang, and Rahul Gupta. 2021. BOLD: Dataset and Metrics for Measuring Biases in Open-Ended Language Generation. (Jan. 2021). arXiv:2101.11718 [cs.CL]

[22] Steven P Dow, Alana Glassco, Jonathan Kass, Melissa Schwarz, Daniel L Schwartz, and Scott R Klemmer. 2011. Parallel prototyping leads to better design results, more divergence, and increased self-efficacy. *ACM Trans. Comput.-Hum. Interact.* 17, 4 (Dec. 2011), 1–24.

[23] Mary T Dzindolet, Scott A Peterson, Regina A Pomranky, Linda G Pierce, and Hall P Beck. 2003. The role of trust in automation reliance. *Int. J. Hum. Comput. Stud.* 58, 6 (June 2003), 697–718.

[24] John Everatt. 1997. The abilities and disabilities associated with adult developmental dyslexia. *J. Res. Read.* 20, 1 (Feb. 1997), 13–21.

[25] Leah Findlater, Steven Goodman, Yuhang Zhao, Shiri Azenkot, and Margot Hanley. 2020. Fairness issues in AI systems that augment sensory abilities. *SIGACCESS Access. Comput.* 125 (March 2020), 1.

[26] Katy Ilonka Gero and Lydia B Chilton. 2019. How a Stylistic, Machine-Generated Thesaurus Impacts a Writer's Process. In *Proceedings of the 2019 on Creativity and Cognition* (San Diego, CA, USA) *(C&C '19)*. Association for Computing Machinery, New York, NY, USA, 597–603.

[27] Katy Ilonka Gero and Lydia B Chilton. 2019. Metaphoria: An Algorithmic Companion for Metaphor Creation. In *Proceedings of the 2019 CHI Conference on Human Factors in Computing Systems* (Glasgow, Scotland Uk) *(CHI '19, Paper 296)*. Association for Computing Machinery, New York, NY, USA, 1–12.

[28] Marjan Ghazvininejad, Xing Shi, Jay Priyadarshi, and Kevin Knight. 2017. Hafez: an Interactive Poetry Generation System. In *Proceedings of ACL 2017, System Demonstrations* (Vancouver, Canada). Association for Computational Linguistics, Stroudsburg, PA, USA.

[29] Dené Granger. 2010. A tribute to my dyslexic body, as I travel in the form of a ghost. *Disabil. Stud. Q.* 30, 2 (April 2010).

[30] Noel Gregg, Chris Coleman, Mark Davis, and Jill C Chalk. 2007. Timed essay writing: implications for high-stakes tests. *J. Learn. Disabil.* 40, 4 (July 2007), 306–318.

[31] Foad Hamidi, Kelly Poneres, Aaron Massey, and Amy Hurst. 2018. Who Should Have Access to My Pointing Data? Privacy Tradeoffs of Adaptive Assistive Technologies. In *Proceedings of the 20th International ACM SIGACCESS Conference on Computers and Accessibility* (Galway, Ireland) *(ASSETS '18)*. Association for Computing Machinery, New York, NY, USA, 203–216. https://doi.org/10.1145/3234695.3239331

[32] IDA Editorial Contributors. 2020. Dyslexia Basics. https://dyslexiaida.org/dyslexia-basics/.

[33] Ji-Youn Jung, Sihang Qiu, Alessandro Bozzon, and Ujwal Gadiraju. 2022. Great Chain of Agents: The Role of Metaphorical Representation of Agents in Conversational Crowdsourcing. In *Proceedings of the 2022 CHI Conference on Human Factors in Computing Systems* (New Orleans, LA, USA) *(CHI '22)*. Association for Computing Machinery, New York, NY, USA, Article 57, 22 pages. https://doi.org/10.1145/3491102.3517653

[34] Shaun K. Kane, Anhong Guo, and Meredith Ringel Morris. 2020. Sense and Accessibility: Understanding People with Physical Disabilities' Experiences with Sensing Systems. In *The 22nd International ACM SIGACCESS Conference on Computers and Accessibility* (Virtual Event, Greece) *(ASSETS '20)*. Association for Computing Machinery, New York, NY, USA, Article 42, 14 pages. https://doi.org/10.1145/3373625.3416990

[35] Anjuli Kannan, Karol Kurach, Sujith Ravi, Tobias Kaufmann, Andrew Tomkins, Balint Miklos, Greg Corrado, Laszlo Lukacs, Marina Ganea, Peter Young, and Vivek Ramavajjala. 2016. Smart Reply: Automated Response Suggestion for Email. In *Proceedings of the 22nd ACM SIGKDD International Conference on Knowledge Discovery and Data Mining* (San Francisco, California, USA) *(KDD '16)*. Association for Computing Machinery, New York, NY, USA, 955–964.

[36] Pranav Khadpe, Ranjay Krishna, Li Fei-Fei, Jeffrey T Hancock, and Michael S Bernstein. 2020. Conceptual Metaphors Impact Perceptions of Human-AI Collaboration. *Proc. ACM Hum.-Comput. Interact.* 4, CSCW2 (Oct. 2020), 1–26.

[37] Alberto Quattrini Li, Licia Sbattella, and Roberto Tedesco. 2013. PoliSpell: An Adaptive Spellchecker and Predictor for People with Dyslexia. , 302–309 pages.

[38] Ryan Louie, Andy Coenen, Cheng Zhi Huang, Michael Terry, and Carrie J Cai. 2020. Novice-AI Music Co-Creation with AI-Steering Tools for Deep Generative Models. In *Proceedings of the 2020 CHI Conference on Human Factors in Computing Systems* (Honolulu, HI, USA) *(CHI '20)*. Association for Computing Machinery, New York, NY, USA, 1–13.

[39] Sônia Maria Pallaoro Moojen, Hosana Alves Gonçalves, Ana Bassôa, Ana Luiza Navas, Graciela de Jou, and Emílio Sánchez Miguel. 2020. Adults with dyslexia: how can they achieve academic success despite impairments in basic reading and writing abilities? The role of text structure sensitivity as a compensatory skill. *Ann. Dyslexia* 70, 1 (April 2020), 115–140.

[40] Meredith Ringel Morris. 2020. AI and accessibility. *Commun. ACM* 63, 6 (May 2020), 35–37.

[41] Meredith Ringel Morris, Adam Fourney, Abdullah Ali, and Laura Vonessen. 2018. Understanding the Needs of Searchers with Dyslexia. In *Proceedings of the 2018 CHI Conference on Human Factors in Computing Systems* (Montreal QC, Canada) *(CHI '18, Paper 35)*. Association for Computing Machinery, New York, NY, USA, 1–12.

[42] Tilly Mortimore and W Ray Crozier. 2006. Dyslexia and difficulties with study skills in higher education. *Studies in Higher Education* 31, 2 (April 2006), 235–251.

[43] Antti Oulasvirta, Janne P Hukkinen, and Barry Schwartz. 2009. When more is less: the paradox of choice in search engine use. In *Proceedings of the 32nd international ACM SIGIR conference on Research and development in information retrieval* (Boston, MA, USA) *(SIGIR '09)*. Association for Computing Machinery, New York, NY, USA, 516–523.

[44] Henriette Folkmann Pedersen, Riccardo Fusaroli, Lene Louise Lauridsen, and Rauno Parrila. 2016. Reading Processes of University Students with Dyslexia - An Examination of the Relationship between Oral Reading and Reading Comprehension. *Dyslexia* 22, 4 (Nov. 2016), 305–321.

[45] Jennifer Pedler. 2007. *Computer Correction of Real-word Spelling Errors in Dyslexic Text.* Ph. D. Dissertation. Birkbeck College, London University.

[46] Marco Pino and Luigina Mortari. 2014. The inclusion of students with dyslexia in higher education: a systematic review using narrative synthesis. *Dyslexia* 20, 4 (Nov. 2014), 346–369.

[47] Geraldine A Price. 2006. Creative solutions to making the technology work: three case studies of dyslexic writers in higher education. *ALT-J* 14, 1 (March 2006), 21–38.

[48] Alec Radford, Jeffrey Wu, Rewon Child, David Luan, Dario Amodei, Ilya Sutskever, and Others. 2019. Language models are unsupervised multitask learners. *OpenAI blog* 1, 8 (2019), 9.

[49] Maria Rauschenberger, Ricardo Baeza-Yates, and Luz Rello. 2019. Technologies for Dyslexia. In *Web Accessibility: A Foundation for Research*, Yeliz Yesilada and Simon Harper (Eds.). Springer London, London, 603–627.

[50] Luz Rello and Ricardo Baeza-Yates. 2016. The Effect of Font Type on Screen Readability by People with Dyslexia. *ACM Trans. Access. Comput.* 8, 4 (May 2016), 1–33.

[51] Luz Rello, Ricardo Baeza-Yates, Stefan Bott, and Horacio Saggion. 2013. Simplify or help? text simplification strategies for people with dyslexia. In *Proceedings of the 10th International Cross-Disciplinary Conference on Web Accessibility* (Rio de Janeiro, Brazil) *(W4A '13, Article 15)*. Association for Computing Machinery, New York, NY, USA, 1–10.

[52] Luz Rello, Miguel Ballesteros, and Jeffrey P Bigham. 2015. A Spellchecker for Dyslexia. In *Proceedings of the 17th International ACM SIGACCESS Conference on Computers & Accessibility* (Lisbon, Portugal) *(ASSETS '15)*. Association for Computing Machinery, New York, NY, USA, 39–47.

[53] Luz Rello and Jeffrey P Bigham. 2017. Good Background Colors for Readers: A Study of People with and without Dyslexia. In *Proceedings of the 19th International ACM SIGACCESS Conference on Computers and Accessibility* (Baltimore, Maryland, USA) *(ASSETS '17)*. Association for Computing Machinery, New York, NY, USA, 72–80.

[54] Luz Rello, Martin Pielot, and Mari-Carmen Marcos. 2016. Make It Big! The Effect of Font Size and Line Spacing on Online Readability. In *Proceedings of the 2016 CHI Conference on Human Factors in Computing Systems* (San Jose, California, USA) *(CHI '16)*. Association for Computing Machinery, New York, NY, USA, 3637–3648.

[55] Lindsay Reynolds and Shaomei Wu. 2018. "I'm never happy with what I write": Challenges and strategies of people with dyslexia on social media. In *Twelfth International AAAI Conference on Web and Social Media.*

[56] Barry Schwartz. 2003. *The Paradox of Choice: Why More Is Less.* Harper Collins.

[57] Ralf Schwarzer and Matthias Jerusalem. 1995. General self-efficacy scale. *Applied Psychology: An International Review* (1995).

[58] Laure M Sharp and Joanne Frankel. 1983. Respondent Burden: A Test of Some Common Assumptions. *Public Opin. Q.* 47, 1 (Jan. 1983), 36–53.

[59] S E Shaywitz, M D Escobar, B A Shaywitz, J M Fletcher, and R Makuch. 1992. Evidence that dyslexia may represent the lower tail of a normal distribution of reading ability. *N. Engl. J. Med.* 326, 3 (Jan. 1992), 145–150.

[60] Romal Thoppilan, Daniel De Freitas, Jamie Hall, Noam Shazeer, Apoorv Kulshreshtha, Heng-Tze Cheng, Alicia Jin, Taylor Bos, Leslie Baker, Yu Du, Yaguang Li, Hongrae Lee, Huaixiu Steven Zheng, Amin Ghafouri, Marcelo Menegali, Yanping Huang, Maxim Krikun, Dmitry Lepikhin, James Qin, Dehao Chen, Yuanzhong Xu, Zhifeng Chen, Adam Roberts, Maarten Bosma, Vincent Zhao, Yanqi Zhou, Chung-Ching Chang, Igor Krivokon, Will Rusch, Marc Pickett, Pranesh Srinivasan, Laichee Man, Kathleen Meier-Hellstern, Meredith Ringel Morris, Tulsee Doshi, Renelito Delos Santos, Toju Duke, Johnny Soraker, Ben Zevenbergen, Vinodkumar Prabhakaran, Mark Diaz, Ben Hutchinson, Kristen Olson, Alejandra Molina, Erin Hoffman-John, Josh Lee, Lora Aroyo, Ravi Rajakumar, Alena Butryna, Matthew Lamm, Viktoriya Kuzmina, Joe Fenton, Aaron Cohen, Rachel Bernstein, Ray Kurzweil, Blaise Aguera-Arcas, Claire Cui, Marian Croak, Ed Chi, and Quoc Le. 2022. LaMDA: Language Models for Dialog Applications. (Jan. 2022). arXiv:2201.08239 [cs.CL]

[61] W Tops, C Callens, E Van Cauwenberghe, J Adriaens, and M Brysbaert. 2013. Beyond spelling: the writing skills of students with dyslexia in higher education. *Read. Writ.* 26, 5 (May 2013), 705–720.

[62] Kristen Vaccaro, Dylan Huang, Motahhare Eslami, Christian Sandvig, Kevin Hamilton, and Karrie Karahalios. 2018. The illusion of control. In *Proceedings of the 2018 CHI Conference on Human Factors in Computing Systems* (Montreal QC Canada). ACM, New York, NY, USA.

[63] Monica van Schaik. 2021. *"ACCEPT THE IDEA THAT NEURODIVERSE KIDS EXIST": DYSLEXIC NARRATIVES AND NEURODIVERSITY PARADIGM VISIONS.* Ph.D. Dissertation. Wilfrid Laurier University.

[64] Ashish Vaswani, Noam Shazeer, Niki Parmar, Jakob Uszkoreit, Llion Jones, Aidan N Gomez, Łukasz Kaiser, and Illia Polosukhin. 2017. Attention is all you need. In *Proceedings of the 31st International Conference on Neural Information Processing Systems* (Long Beach, California, USA) *(NIPS'17)*. Curran Associates Inc., Red Hook, NY, USA, 6000–6010.

[65] Laura Weidinger, John Mellor, Maribeth Rauh, Conor Griffin, Jonathan Uesato, Po-Sen Huang, Myra Cheng, Mia Glaese, Borja Balle, Atoosa Kasirzadeh, Zac Kenton, Sasha Brown, Will Hawkins, Tom Stepleton, Courtney Biles, Abeba Birhane, Julia Haas, Laura Rimell, Lisa Anne Hendricks, William Isaac, Sean Legassick, Geoffrey Irving, and Iason Gabriel. 2021. Ethical and social risks of harm from Language Models. (Dec. 2021). arXiv:2112.04359 [cs.CL]

[66] Shaomei Wu, Lindsay Reynolds, Xian Li, and Francisco Guzmán. 2019. Design and Evaluation of a Social Media Writing Support Tool for People with Dyslexia. In *Proceedings of the 2019 CHI Conference on Human Factors in Computing Systems* (Glasgow, Scotland Uk) *(CHI '19, Paper 516)*. Association for Computing Machinery, New York, NY, USA, 1–14.

[67] Tongshuang Wu, Ellen Jiang, Aaron Donsbach, Jeff Gray, Alejandra Molina, Michael Terry, and Carrie J Cai. 2022. PromptChainer: Chaining Large Language Model Prompts through Visual Programming. (March 2022). arXiv:2203.06566 [cs.HC]

[68] Jing Xu, Arthur Szlam, and Jason Weston. 2021. Beyond Goldfish Memory: Long-Term Open-Domain Conversation. (July 2021). arXiv:2107.07567 [cs.CL]

[69] Ann Yuan, Andy Coenen, Emily Reif, and Daphne Ippolito. 2022. Wordcraft: Story Writing With Large Language Models. In *27th International Conference on Intelligent User Interfaces* (Helsinki, Finland) *(IUI '22)*. Association for Computing Machinery, New York, NY, USA, 841–852.

[70] Wojciech Zaremba, Greg Brockman, and OpenAI. 2021. OpenAI Codex. https://openai.com/blog/openai-codex/.

[71] Tony Z Zhao, Eric Wallace, Shi Feng, Dan Klein, and Sameer Singh. 2021. Calibrate Before Use: Improving Few-Shot Performance of Language Models. (Feb. 2021). arXiv:2102.09690 [cs.CL]